# Environmental Sound Classification on the Edge: A Pipeline for Deep Acoustic Networks on Extremely Resource-Constrained Devices

Md Mohaimenuzzaman[*], Christoph Bergmeir, Ian West[1], Bernd Meyer

*Department of Data Science and AI, Monash University, Wellington Rd, Clayton, 3800, VIC, Australia*

ABSTRACT

Significant efforts are being invested to bring state-of-the-art classification and recognition to edge devices with extreme resource constraints (memory, speed, and lack of GPU support). Here, we demonstrate the first deep network for acoustic recognition that is small, flexible and compression-friendly yet achieves state-of-the-art performance for raw audio classification. Rather than handcrafting a once-off solution, we present a generic pipeline that automatically converts a large deep convolutional network via compression and quantization into a network for resource-impoverished edge devices. After introducing ACDNet, which produces above state-of-the-art accuracy on ESC-10 (96.65%), ESC-50 (87.10%), UrbanSound8K (84.45%) and AudioEvent (92.57%), we describe the compression pipeline and show that it allows us to achieve 97.22% size reduction and 97.28% FLOP reduction while maintaining close to state-of-the-art accuracy 96.25%, 83.65%, 78.27% and 89.69% on these datasets. We describe a successful implementation on a standard off-the-shelf microcontroller and, beyond laboratory benchmarks, report successful tests on real-world datasets.

## 1. Introduction

Intelligent sound recognition is receiving strong interest in a growing number of application areas, from technical safety [1–3], surveillance and urban monitoring [4–6] to environmental applications [7,8].

We are specifically interested in acoustic monitoring of animal vocalisations, a useful and well-established methodology in biodiversity management [9–12]. Traditionally, animal monitoring is conducted using passive acoustic recording and subsequent manual evaluation by human experts. Since this is extremely labour-intensive, AI-based solutions have recently been at the center of interest [9].

The fact that animal monitoring often has to take place in remote regions [10] poses some interesting technical challenges for automated audio recognition. Almost all suitable state-of-the-art (SOTA) methods for environmental sound classification (ESC) are based on Deep Learning (DL) [13–16] and consequentially have very high computational requirements. Current AI-based animal monitoring systems typically require recordings to be uploaded to cloud-based platforms or high-powered desktop machines that have sufficient resources to process AI audio models [11,17]. However, remote areas often have insufficient network coverage or no network coverage at all, rendering the upload of audio data impossible. Continuous long-term monitoring, which is essential for biodiversity management [10], can thus only become a reality if recognition can take place directly on the monitoring devices in the field.

There are different DL-based pipelines/frameworks proposed for the fields like computer vision [18–21] and video analysis [22]; however, no equivalent method is proposed for sound classification.

Our goal thus is to make acoustic classification possible on small, embedded edge devices, which generally have very little memory and compute power owing to tight constraints on power consumption.

While the reasons to move the recognition directly onto small edge devices are particularly compelling in the case of ecological monitoring, similar considerations apply to many other applications of intelligent sound recognition, from industrial safety to consumer devices, for a variety of reasons, including cost and convenience. Specifically cost is a factor that commonly limits the available compute power and network bandwidth in edge applications.

Significant efforts have already been invested in developing smaller, more efficient convolutional neural network (CNN) models for mobile devices [23,24], smartphones, CPUs, and GPUs [25–27].

---

[*] Corresponding author.
   *E-mail addresses:* md.mohaimen@monash.edu (M. Mohaimenuzzaman), christoph.bergmeir@monash.edu (C. Bergmeir), ian.west@monash.edu (I. West), bernd.meyer@monash.edu (B. Meyer).
[1] At the time this research work was conducted

Table 1
SOTA models for ESC-10, ESC-50, US8K and AE datasets sorted by year of publication. **Abbreviations**: ATTN (Attention), CO (Cochleagram), CRP (Cross Recurrence Plot), CT (Chromagram), DGT (The Discrete Gabor Transform), ENS (Ensemble Model), FBE (FilterBank Energies), GT (GammaTone), LP (Log-Power Spectrogram), Mel (Mel Spectrogram), MFCC (Mel-Frequency Cepstral Coeficients), PE (Phase-Encoded), Raw (Raw audio wave), Spec (Spectrogram), TEO (Teagerfis Energy Operator).

| Networks | #Channels | Input(s) | Accuracy (%) on Datasets | | | |
|---|---|---|---|---|---|---|
| | | | ESC-10 | ESC-50 | US8K | AE |
| Human [38] | - | - | 95.70 | 81.30 | - | - |
| Piczak-CNN [47] | Multi | Mel | 90.20 | 64.50 | 73.70 | - |
| GSTC⊕TEO-GSTC [48] | Multi | TEO, GT | - | 81.95 | - | - |
| GTSC⊕ConvRBM [40] | Multi | (PE)FBE | - | 83.00 | - | - |
| FBEs⊕PEFBEs [49] | Multi | FBE | - | 84.15 | - | - |
| EnvNet [50] | Single | Raw | 88.10 | 74.10 | 71.10 | - |
| EnvNet-v2 [43] | Single | Raw | 88.80 | 81.60 | 76.60 | - |
| EnvNet-v2 + BC [43] | Single | Raw | 91.30 | 84.70 | 78.30 | - |
| GoogLENet [51] | Multi | Mel, MFCC, CRP | 86.00 | 73.00 | 93.00 | - |
| Kumar-CNN [52] | Multi | Mel | - | 83.50 | - | - |
| VGG-CNN+Mixup [53] | Multi | Mel, GT | 91.70 | 83.90 | 83.70 | - |
| AclNet (WM = 1.5) [44] | Single | Raw | - | 85.65 | - | - |
| TSCNN-DS [15] | ENS, Multi | Mel, MFCC, CT | - | - | 97.20 | - |
| Multi-stream [54] | Multi, ATTN | Spec | 94.20 | 84.00 | - | - |
| CRNN [55] | Multi+ATTN | Spec | - | 85.20 | - | - |
| FBEs⊕ConvRBM [40] | Multi | Spec | - | 86.50 | - | - |
| ESResNet [14] | Multi | LP | **97.00** | 91.50 | 85.42 | - |
| Multi-CNN [41] | Multi | Mel | - | 89.50 | - | - |
| WEANET [13] | Multi | Spec | - | **94.10** | - | - |
| CNN [42] | ENS, Multi | DGT, Mel, GT, CO | - | 88.65 | - | - |
| BNN-GAP8 [56] | Multi | Spec | - | - | - | 77.90 |
| CNN-CNP [57] | Multi | Spec | - | - | - | 85.10 |
| Method B [16] | Multi | Spec | - | - | - | **92.80** |

However, smartphones and similar devices still have computing power orders of magnitude above the embedded devices we are targeting and thus, much more radical minimisation is required.

Embedded edge devices use energy efficient microcontroller units (MCUs) and are typically based on system-on-a-chip (SoC) hardware with less than 512kB of RAM and slow clock speeds. GPU support is available on comparatively high-powered specialised edge devices, such as Google's Coral TPU and Nvidia's Jetson, but not on the ultra low-power MCUs we are targeting. The basis of some of the most common energy efficient SoCs, like the Nordic nRF52840 [28] and the STM32F4 [29] is the 32 bit ARM Cortex M4F CPU, typically run at clock speeds below 200 MHz.

To the best of our knowledge, practical acoustic classification has not previously been achieved on such extremely small devices beyond very simple tasks, such as wake-word recognition. Deploying DL models on such MCUs requires aggressive minimisation techniques like model compression [30–33], knowledge distillation [34] and quantization [30,35]. Some works have used such techniques for efficient models in computer vision [36,37], but acoustic classification has not benefited from this yet.

We present an acoustic classification solution for energy-efficient edge devices that achieves close to SOTA performance for raw audio classification on ESC-10&50 [38], UrbanSound8K (US8K) [39] and AudioEvent (AE) [16], the benchmarks that are used to assess large, non resource-constrained networks. Importantly, we do not specifically design the network for these target devices. Rather, we describe a generic pipeline that automatically compress and quantize a large deep CNN into a network suitable for edge devices.

We first introduce ACDNet, a new, flexible and compression friendly sound classification architecture that exceeds current SOTA performance on raw audio classification with accuracies of 96.65±0.06%, 87.10±0.02%, 84.45±0.05% and 92.57±0.05% on ESC-10, ESC-50, US8K and AE, respectively. These performances are also very close to the overall SOTA for the mentioned benchmarks (see Table 1). We then compress ACDNet using a novel network-independent compression approach to obtain an extremely small model (Micro-ACDNet). Despite 97.22% size reduction and 97.28% reduction in FLOPs, Micro-ACDNet still achieves classification accuracy of 96.25%, 83.65%, 78.28% and 89.69% on the same datasets, which are significantly higher than human accuracy (i.e., 81.30% for ESC-50) and still very close to the SOTAs.

To classify 1.5s audio sampled at 20kHz, ACDNet uses 4.74M parameters (18.06MB) and requires approximately 544M FLOPs for a single inference. On the other hand, the micro version, Micro-ACDNet has only 0.131M parameters (0.50MB) and requires only 14.82M FLOPs for an inference, which is well within the capabilities of the MCUs we are targeting. We describe a successful deployment of a model for classifying 50 classes on a standard off-the-shelf MCU and, beyond laboratory benchmarks, report successful tests on real-world data.

To the best of our knowledge, this is the first time a deep network for sound classification of 50 classes has successfully been deployed on an edge device. While this should be of interest in its own right, we believe it to be of particular importance that this has been achieved with a generic conversion pipeline rather than hand-crafting a network for minimal size.

Hence, the contributions of this research article are as follows: 1) it introduces the first generic pipeline for sound classification using DL in extremely resource-constrained MCUs; 2) it presents ACDNet, a flexible and compression-friendly SOTA DL model architecture for raw audio classification; 3) instead of hand-crafting a once-off solution, it introduces a novel deep CNN compression approach to obtain tiny models for MCUs; and 4) finally, after 8-bit quantization, we deploy Micro-ACDNet on a standard othe shelf MCU with successful tests on real-world data.

## 2. Related work

A wide range of DL models for acoustic classification achieve SOTA performance on different environmental sound classification benchmarks. Table 1 shows the SOTA models developed over the past years for the four datasets discussed (ESC-10, ESC-50, US8K and AE). Unfortunately, most of them have not adequately disclosed their model sizes and computation requirements. In reality, none of the latest SOTA models for any of the four datasets [13,14,40–42] has provided information on model sizes and computation requirements. EnvNet-v2 [43] and AclNet [44] are found to be the only ones of the latest twenty two SOTA models presented in Table 1 that have fully detailed their computational requirements (see Table 6). The high requirements of EnvNet-v2 are partly due to the use of multiple dense layers [43], which, regrettably, do not lend themselves well to compression [45,46].

**Table 2**
ACDNet architecture. Output shape represents (channel, frequency, time), $i\_len$ is the input length, $n\_cls$ is the number of output classes, $sr$ is the sampling rate in Hz.

| Layers | Kernel Size | Stride | Filters | Output Shape | Block |
|---|---|---|---|---|---|
| Input | | | | $(1, 1, w = i\_len)$ | |
| conv1 | (1, 9) | (1, 2) | $x$ | $(x, 1, w = \frac{w-9}{2} + 1)$ | SFEB |
| conv2 | (1, 5) | (1, 2) | $x\prime = x * 2^3$ | $(x\prime, 1, w = \frac{w-5}{2} + 1)$ | |
| Maxpool1 | (1, SFEB_PS) | (1, SFEB_PS) | | $(x\prime, 1, w = \frac{w}{ps})$ | |
| swapaxes | | | | $(1, h = x\prime, w)$ | |
| conv3 | (3, 3) | (1,1) | $x\prime = x * 2^2$ | $(x\prime, h, w)$ | TFEB |
| Maxpool2 | TFEB_PS[0] | TFEB_PS[0] | | $(x\prime, h = \frac{h}{kh}, w = \frac{w}{kw})$ | |
| conv4,5 | (3, 3) | (1, 1) | $x\prime = x * 2^3$ | $(x\prime, h, w)$ | |
| Maxpool3 | TFEB_PS[1] | TFEB_PS[1] | | $(x\prime, h = \frac{h}{kh}, w = \frac{w}{kw})$ | |
| conv6,7 | (3, 3) | (1, 1) | $x\prime = x * 2^4$ | $(x\prime, h, w)$ | |
| Maxpool4 | TFEB_PS[2] | TFEB_PS[2] | | $(x\prime, h = \frac{h}{kh}, w = \frac{w}{kw})$ | |
| conv8,9 | (3, 3) | (1, 1) | $x\prime = x * 2^5$ | $(x\prime, h, w)$ | |
| Maxpool5 | TFEB_PS[3] | TFEB_PS[3] | | $(x\prime, h = \frac{h}{kh}, w = \frac{w}{kw})$ | |
| conv10,11 | (3, 3) | (1, 1) | $x\prime = x * 2^6$ | $(x\prime, h, w)$ | |
| Maxpool6 | TFEB_PS[4] | TFEB_PS[4] | | $(x\prime, h = \frac{h}{kh}, w = \frac{w}{kw})$ | |
| Dropout (0.2) | | | | | |
| conv12 | (1, 1) | (1, 1) | $n\_cls$ | $(n\_cls, h, w)$ | |
| Avgpool1 | TFEB_PS[5] | TFEB_PS[5] | | $(n\_cls, 1, 1)$ | |
| Flatten | | | | $(n\_cls)$ | |
| Dense1 | | | $n\_cls$ | $(n\_cls)$ | |
| Softmax | | | | $(n\_cls)$ | Output |

Table 1 shows that all the recent SOTA approaches (sorted by year of publication) resort to multi-channel image-like spectrogram-based inputs, multiple parallel blocks or ensembles models and attention mechanisms. While this demonstrates some success in improving performance, it also increases the input and model sizes significantly and thus does not constitute an advantageous starting point for our purposes. The whole SOTA list also shows no spectrogram-based SOTA model to date runs on single-channel input. The reason behind this lack of success in inputting audio as grey images (single-channel spectrogram) is that unlike vertical and horizontal axes of images, the axes (time & frequency) of spectrograms are not homogeneous and hence do not carry the same spatial information [5,58,59]. Using multiple channels to input different feature sets usually compensates for this problem; however, this merely makes the input size bigger than the available primary memory available in a typical MCU. On the other hand, inputting audio as raw waveform does not require expensive hand-designed features, allows us to exploit better modelling capability and learning representations [5] yet is suitable for single-channel input for MCUs. Table 1 also shows that all the SOTA models running on single-channel input runs on raw audio waveform. Considering all the above discussed issues, this research works with DL models that run on single-channel raw audio waveform.

Some recent work in computer vision that has specifically focused on very small CNN models is highly relevant to our work. MCUNet [37] starts from SOTA image classification models, such as ResNet50 [60] and MobileNetV2 [23], and uses search-based optimisation to find models that fit on MCUs. The authors report >70% accuracy after deploying the final models on MCUs.

The recent Sparse Architecture Search (SpArSe [36]) appears to be a very promising approach. It uses multi-objective search to automatically construct models small enough for MCU deployment. It optimises accuracy, model size, and working memory size by performing a search over pruning, parameters, and model architecture. The latter fundamentally sets it apart from other work. While other work minimises specific target models, SpArSe includes the model architecture in the search space. In this way, SpArSe achieves impressive results with models that achieve high accuracy on a variety of standard vision benchmarks (MNIST, CIFAR10, CUReT, Chars4k) and are small enough to be deployed on standard MCUs.

Model search can require prohibitively vast amounts of computation time. The feasibility of SpArSe is achieved by constraining the search space to model proposals that are specific morphs of a defined starting point. The success of SpArSe thus relies on the availability of suitable models as starting points. While still improving their performance and size, [36] does in fact start from models which themselves are already constructed to fit on MCUs (Bonsai [61] and ProtoNN [62]). Such starting points are currently not available for audio classification, and, to the best of the authors knowledge, no comparable work exists for this problem domain.

In the audio domain, the only work that specifically has the reduction of computational requirements as its primary focus is Edge-L$^3$ [63]. It achieves a model size of 0.814MB by pruning L$^3$-Net [64], which has an initial size of 18MB. While this approach achieves good theoretical compression ratios and good prediction accuracy, it relies on unstructured compression. This results in sparse matrix models which, unfortunately, do not ideally lend themselves to a direct implementation on embedded devices. MCUs generally lack the dedicated hardware and software support for sparse matrix computations required to capitalise on these theoretical savings [45,46]. Hence, structured compression techniques that produce dense matrix models promise to be a preferable approach for targeting MCUs.

In the following sections, we detail the construction of a flexible and compression friendly DL model that betters current SOTA performance for raw audio classification on ESC-10, ESC-50, US8K and AE while also providing a suitable starting point for compression, followed by a discussion of the structured compression methods used to minimise this for MCU deployment.

## 3. Proposed base network

We present a new model architecture (ACDNet) for acoustic classification that is smaller, flexible, compression-friendly and more efficient than the current SOTA networks, yet produces classification accuracy close to them and exceeds classification accuracy of the SOTA models for raw audio classification on the most widely used Environmental Sound Classification datasets ESC-10, ESC-50, US8K and AE.

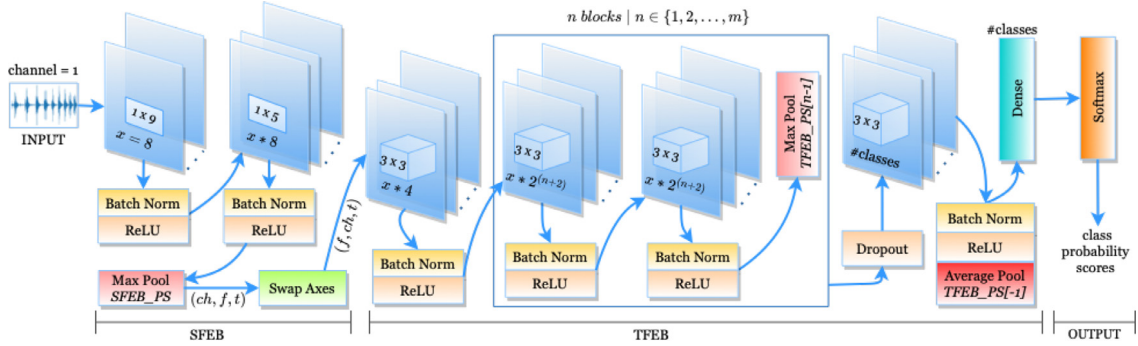

**Fig. 1.** Visual representation of ACDNet model architecture.

### 3.1. ACDNet Architecture

Unlike other suggested SOTA models, which use pre-extracted features and multi-channel input (such as hand-crafted features and spectrograms), the ACDNet architecture concentrates solely on feature extraction through convolution layers. All the convolution layers are followed by a batch normalization and a ReLU activation layers. The maxpool layers have strides equal to their pool size to avoid overlapping. The network is fed with raw audio time series (i.e., single-channel input). ACDNet consists of two feature extraction blocks followed by an output block. The feature extraction blocks are Spectral Feature Extraction Block (SFEB) and Temporal Feature Extraction Block (TFEB). Fig. 1 depicts the layer structure of different blocks of the ACDNet architecture.

SFEB consists of two 1-D strided convolutions followed by the first pooling layer. This block extracts low level audio features (spectral features) from raw audio through convolutions at a frame rate of 10ms. The output of this block is achieved by downsampling the convolution output using the maxpool with kernel size determined by the following equation:

$$SFEB\_PS = \frac{w}{((i\_len/sr) * 1000)/10} \quad (1)$$

where $w$ is the width of the output of previous layer, $i\_len$ is the input length and $sr$ is the sampling rate.

The axes of the data produced by SFEB are swapped from (ch, f, t) to (ch'=f, f'=ch, t'=t), and the result is fed as input to TFEB to convolve over both frequency and time for extracting high level features also known as hierarchical temporal features. Thus, this block of the network works like a convolution on images. TFEB consists of convolutions 3–12 in Table 2. The first convolution layer is followed by a pooling layer. After that, the convolution layers (4–11) are stacked like VGG-13 [65] (two convolutions followed by a pooling layer). The final convolution layer (conv-12) is followed by a single average pooling layer. The kernel sizes of the pooling layers are determined by $TFEB\_PS = \{(f(x_h, i), f(x_w, i))_k\}_{k \in \{1,2,...,N\}}$ where $x_h$ and $x_w$ are the height and width of the input to TFEB block, $i$ is the index of the pooling layer, $N$ is the number of pooling layers, and $f(x, i)$ is defined by:

$$f(x, i) = \begin{cases} 2 & \text{if } x > 2 \text{ and } i < N \\ 1 & \text{if } x = 1 \\ \frac{x}{2^{(N-1)}} & \text{if } i = N \end{cases} \quad (2)$$

TFEB block ends with an average-pool layer followed by a dense layer. The dense layer have output neurons equal to the number of classes to make sure the size of the output vector of TFEB is always equal to the number of classes. This dense layer is crucial for the compression of the network in a later stage. The output of the TFEB is fed to a softmax output layer for classification. Table 2 shows the overall ACDNet architecture.

While the structure of the first two convolution layers has some similarity with EnvNet-v2 [43] and the other convolution layers with AclNet [44], EnvNet-v2 and AclNet architectures cannot flexibly be adapted to different lengths of audio recorded at different sampling rates. Furthermore, they are not compression friendly for deriving tiny network models suitable for MCUs with comparable accuracy.

EnvNet-v2 has eight 1D convolutions and two 2D convolutions with 1056 filters, followed by two massive dense layers with 4096 neurons each. These dense layers produce approximately 100M parameters out of 101M These dense layers produce approximately 100M parameters out of 101M parameters of the entire network. Hence, these two layers will be heavily pruned during the compression process, which would heavily impact the network's performance.

On the other hand, AclNet has twelve convolution layers with 3050 filters followed by the softmax output layer. The number of output filters from the last convolution layer always equals to the number of classes. Hence, this last convolution layer cannot be compressed. This might lead to pruning some more important filters from other layers, which may cause further accuracy loss. In addition to that, this may make the network structure pyramid-shaped which is wider towards the end when the number of classes is higher (e.g., 50 or 100). Such wide layers require extra primary memory during inference time that a typical MCU would struggle to provide.

ACDNet architecture is strongly motivated from the perspective of compressibility of the network and adaptability with respect to different audio lengths recorded at different sampling rates (see Table 2). It addresses the issue of EnvNet-v2 by adding more convolution layers and a tiny dense layer followed by the softmax output layer to allow the network to adapt to the changes in data shape when model compression is applied. The addition of a tiny dense layer also addresses AclNet's compressibility issue.

### 3.2. Experimental setup

ACDNet is implemented in PyTorch version 1.7.1 and *Wavio* audio library is used to process the audio files. The full code is available at: https://github.com/mohaimenz/acdnet

#### 3.2.1. Datasets

The experiments are conducted on three of the most robust and widely used audio benchmark datasets - Environmental Sound (ESC-10, ESC-50 [38] and US8K [39]) and another audio dataset named AudioEvent(AE) [16].

ESC-50 contains 2000 samples (5-sec-long audio recordings, sampled at 16kHz and 44.1kHz) which are equally distributed over 50 balanced disjoint classes (40 audio samples for each class). Furthermore, a division of the dataset into five splits is available, helping researchers to achieve unbiased comparable results in 5-fold cross validation (CV). ESC-10 is a subset of ESC-50 with ten classes and 400 samples equally distributed over the classes. US8K dataset contains 8732 labelled audio clips (<=4s) of urban sounds from 10 classes recorded at 22.05kHz. The clips are pre-sorted to 10 folds for 10-fold CV to achieve unbiased comparable results. The AE dataset has 5223 samples unevenly distributed over 28 classes recorded at 16kHz.

Our work is ultimately aimed at real-world applications for biodiversity, and these can be more difficult than standard benchmarks, as evidenced in the LifeCLEF competition [66]. However, LifeCLEF itself is not suitable as a benchmark for the problem tackled here as this starts from strong labeling, whereas LifeCLEF explores weak labeling. As a real-world test, we evaluate ACDNet on a dataset of frog recordings that is known to be challenging for conventional animal monitoring solutions [67,68]. It contains 9132 field recordings of 10 different frog species across a variety of locations in Australia sampled at 32kHz.

#### 3.2.2. Data processing and training setup

For setting up the output filters of the convolutions of ACDNet, we set $x = 8$ to the *conv*1 layer of the *SFEB* block (see Table 2). The network is trained and tested on 20kHz data with inputs of length 30,225 (approximately 1.51s audio). We downsampled the data to 20kHz in order to reduce input size, model size, and power consumption. We have not observed any difference in performance with audios re-sampled at a lower sampling rate.

We follow the data preprocessing, augmentation, and mixing of classes described in EnvNet-v2 [43] to produce training samples. According to EnvNet-v2, two training sounds belonging to two different classes are randomly picked, padded with T/2 (T = input length) zeros to each side of both the samples and a *T-s* section from both the sounds is randomly cropped. Then, the two cropped samples are mixed using a random ratio. We denote the maximum gains of the cropped samples $s_1$, $s_2$ by $g_1$, $g_2$, and $r$ is the random ratio between (0,1). The ratio of the mixed sounds $p$ according to EnvNet-v2, is

$$p = \frac{1}{1 + 10 * \frac{g_1 - g_2}{20} * \frac{1-r}{r}} \quad (3)$$

Finally, the mixed sound sample $S_{mix}$ for training is determined by

$$S_{mix} = \frac{ps_1 + (1-p)s_2}{\sqrt{p^2 + (1-p)^2}} \quad (4)$$

In the testing phase, we pad T/2 zeros to each side of the test input sample and then extract ten windows (each of length 30,225) at a regular interval of T/9 as input for the network. The input data for training and testing are normalized by dividing them by 32,768, the full range of 16-bit recordings.

The network is trained for 2000 epochs with an initial learning rate of 0.1 along with a learning rate scheduler {600, 1200, 1800} decaying at a factor of 10. The first 10 epochs are considered as warm-up epochs and a 0.1 times smaller learning rate is used for these 10 epochs. Since we mix up class labels to generate samples, the mini-batch ratio labels should represent the expected class probability distribution. Hence, we use *KLDivLoss* (Kullback-Leibler Divergence Loss) as the loss function instead of cross-entropy loss [43]. We optimize it using back-propagation and Stochastic Gradient Descent (*SGD*) with a *Nesterov momentum* of 0.9, a weight decay of 5e-4, and a batch size of 64. Furthermore, the weights of all convolution and dense layers are initialized using the *he_normal* [69] initialization method. At the end of the training and validation, the best model is used as the final model. The loss function optimized is shown in Eq. 5. Here, $x$ is the input minibatch, $f_\theta(x)$ is the approximation and $y$ is the true distribution of labels for the input data. Furthermore, $n$ is the mini-batch size, $m$ is the number of classes, $\eta$ is the learning rate, and $\theta \leftarrow \theta - \eta \frac{\partial L}{\partial \theta}$.

**Table 3**
CV Accuracy and estimated 95% CI of classification accuracy of ACDNet on ESC-10, ESC-50, US8K and AE datasets.

| Datasets | Classification Accuracy (%) | |
|---|---|---|
| | CV | 95-PCI |
| ESC-10 | 96.65±0.06 | 96.73±0.04 |
| ESC-50 | 87.10±0.02 | 87.15±0.06 |
| US8K | 84.45±0.05 | 84.34±0.03 |
| AE | 92.57±0.05 | 92.56±0.04 |

$$L = \frac{1}{n} \sum_{i=1}^{n} (D_{KL}(y^{(i)} || f_\theta(x^{(i)}))) = \frac{1}{n} \sum_{i=1}^{n} \sum_{j=1}^{m} y_j^{(i)} \log \frac{y_j^{(i)}}{(f_\theta(x^{(i)}))_j} \quad (5)$$

#### 3.2.3. Training and testing ACDNet

We conduct 5-fold CV for ESC-10&50, 10-fold CV for US8K and 80-20 train-test split for AE over 10 independent runs.

To establish the statistical significance of comparisons between ACDNet and other SOTA models, we cannot apply typical standard procedures, such as the Friedman test with post-hoc analysis or the ROC curve. This is because the implementations of the comparison models are not available, and we thus have to rely on published data only. Generally, only the mean accuracy is reported (without standard error), and only a few models have been tested on all four datasets. Instead, we use the bootstrap confidence intervals [70,71] to calculate the 95 percent confidence interval (95-PCI) [72] of classification accuracy of ACDNet for all four datasets to confirm the model's generalizability and reliability.

We run ACDNet 1000 times on the test sets of all four datasets using bootstrap sampling with replacement, and then construct the 95-PCI of classification accuracy using the following formula.

$$CI = \mu \pm Z \frac{\sigma}{\sqrt{N}} \quad (6)$$

where $\mu$ is the mean accuracy of all the tests, $Z = 1.96$ [72], $\sigma$ is the standard deviation and $N$ is the number of tests. Table 3 presents the CV accuracy and the estimated 95-PCI of classification accuracy of ACDNet on the mentioned datasets.

Table 3 reveals that the CV accuracy of ACDNet and its estimated 95-PCI accuracy on various datasets are almost equivalent, if not identical, indicating that the network is well generalized over all the datasets.

We now add ACDNet to the SOTA table for raw audio classification. Table 4 indicates that ACDNet outperforms the current SOTA with an overall accuracy of 96.65 ± 0.06% on ESC-10, 87.10 ± 0.02% on ESC-50, 84.45 ± 0.05% on US8K and 92.57 ± 0.05% on AE datasets. Furthermore, when compared to all SOTA models regardless of the types of the input they deal with, ACDNet achieves comparable prediction accuracy (see Table 5).

For the real-world frog dataset, ACDNet is trained for 1000 epochs. We obtain 5-fold CV accuracy of 88.98% even without data augmentation.

Few SOTA models report their size and computation requirements in the literature. According to our knowledge on current literature, EnvNet-v2 and AclNet are the only two of the top-ten models for either of the two datasets to report parameter count, size, and FLOPs. Table 6 shows that ACDNet requires 21.39× less memory and 2.98× less FLOPs than EnvNet-v2 and 2.25× less memory and 1.97× less FLOPs than AclNet, respectively.

**Table 4**
ACDNet in the SOTA leaderboard for raw audio classification on ESC-10&50, US8K and AE datasets. Accuracy values with asterisk (*) are reproduced by us.

| Networks | Accuracy (%) on Datasets | | | |
|---|---|---|---|---|
| | ESC-10 | ESC-50 | US8K | AE |
| EnvNet [50] | 88.10 | 74.10 | 71.10 | - |
| EnvNet-v2 [43] | 88.80 | 81.60 | 76.60 | - |
| EnvNet-v2 + BC [43] | 91.30 | 84.70 | 78.30 | - |
| AclNet [44] | 95.75* | 85.65 | 79.17* | 90.15* |
| **ACDNet (ours)** | **96.65 ± 0.06** | **87.10 ± 0.02** | **84.45 ± 0.05** | **92.57 ± 0.05** |

**Table 5**
ACDNet in overall SOTA leaderboard of ESC-10&50, US8K and AE datasets. Column 'Acc' presents model accuracy. Accuracy values with asterisk (*) are reproduced by us.

| Datasets → | ESC-10 | | ESC-50 | | US8K | | AE | |
|---|---|---|---|---|---|---|---|---|
| Networks ↓ | Acc (%) | Rank | Acc (%) | Rank | Acc (%) | Rank | Acc (%) | Rank |
| Piczak-CNN [47] | 90.20 | 7 | 64.50 | 13 | 73.70 | 9 | - | - |
| GoogLENet [51] | 86.00 | 10 | 73.00 | 12 | 93.00 | 2 | - | - |
| EnvNet [50] | 88.10 | 9 | 74.10 | 11 | 71.10 | 10 | - | - |
| EnvNet-v2 [43] | 88.80 | 8 | 81.60 | 10 | 76.60 | 8 | - | - |
| EnvNet-v2 + BC [43] | 91.30 | 6 | 84.70 | 7 | 78.30 | 7 | - | - |
| VGG-CNN+Mixup [53] | 91.70 | 5 | 83.90 | 9 | 83.70 | 5 | - | - |
| Multi-stream+Attn[54]. | 94.20 | 4 | 84.00 | 8 | - | - | - | - |
| AclNet (WM = 1.5) [44] | 95.75* | 3 | 85.65 | 6 | 79.17* | 6 | 90.15* | 3 |
| FBEs⊕ConvRBM [40] | - | - | 86.50 | 5 | - | - | - | - |
| CNN [42] | - | - | 88.65 | 3 | - | - | - | - |
| ESResNet [14] | 97.00 | 1 | 91.50 | 2 | 85.42 | 3 | - | - |
| TSCNN-DS [15] | - | - | - | - | 97.20 | 1 | - | - |
| WEANET [13] | - | - | 94.10 | 1 | - | - | - | - |
| BNN-GAP8 [56] | - | - | - | - | - | - | 77.90 | 5 |
| CNN-CNP [57] | - | - | - | - | - | - | 85.10 | 4 |
| Method B [16] | - | - | - | - | - | - | 92.80 | 1 |
| **ACDNet (ours)** | **96.65** | **2** | **87.10** | **4** | **84.45** | **4** | **92.57** | **2** |

**Table 6**
Parameters, size, and computation requirements for current SOTA models on the ESC-50 dataset. Here, $x$ denotes the size and FLOPs of ACDNet, for the Size and FLOPs column, respectively.

| Networks | #Filters | Params(M) | Size(MB) | FLOPs(M) |
|---|---|---|---|---|
| EnvNet-v2 [43] | 1056 | 101.25 | 386.25=21.39$x$ | 1620=2.98$x$ |
| AclNet [44] | 3050 | 10.63 | 40.57=2.25$x$ | 1070=1.97$x$ |
| **ACDNet (ours)** | 2074 | 4.74 | 18.06=$x$ | 544=$x$ |

## 4. Network compression

As discussed before, unstructured compression is not suitable for MCUs. Therefore, we introduce a new class of structured hybrid compression techniques. We first sparsify the weight matrices of ACDNet using unstructured compression (e.g., L0-Norm) and then apply structured compression. The idea of sparsifying the weight matrices is inspired by the fact that bringing sparsity into the network reduces the chance of model over-fitting and enhances the model performance [73]. Furthermore, many researchers have shown that sparse pruning of neural networks often produces the same or even better accuracy than the base network [30,74,75]. This allows us to focus the structured compression on the important weights.

There are different techniques by which structured compression can be achieved, such as Channel Pruning [32], Weight Sharing [30,76], Huffman Coding [77], Knowledge Distillation [34,78] and Quantization [30,79]. Though many works in the literature use these techniques in computer vision, they have hardly ever been used for audio tasks. Due to this lack of guidance from the literature, we use the best established pruning-based model compression techniques proposed for computer vision to compress ACDNet. Furthermore, since quantization does not conflict with other compression techniques, we use quantization of the compressed model to further reduce its size before deploying it on the embedded device.

Structured pruning, namely channel pruning for CNN is conducted by ranking the channels globally and removing the lowest ranked channels. Let $z_l$ be the feature maps of a network with $z_l \in \mathbb{R}^{h_l \times w_l \times c_l}$ where $h_l \times w_l$ is the dimension and $c_l$ are the channels of layers with $l \in [1, 2, \ldots, L]$. $z_l^{(i)}$ is an individual feature map with $i \in [1, 2, \ldots, c_l]$. Thus, the ranking is calculated (Eq. 8) followed by $\Omega$, the layer-wise normalization (Eq. 7) of the feature maps. Finally, the pruning candidate is determined by Eq. 9:

$$\Omega(z_l) = \frac{|z_l^{(i)}|}{\sqrt{\sum |z_l|^2}} \quad (7)$$

$$z_{rank} = sort(\Omega(z_l)) \quad (8)$$

$$z_{prune} = argmin(z_{rank}) \quad (9)$$

Many ranking criteria exist in the current literature. Few of them are magnitude-based (L2-Norm) ranking [30,33], Taylor criteria-based ranking [32] and binary index-based ranking in AutoPruner [80]. In this paper, we explore the magnitude-based ranking which is one of the very common techniques and the Taylor criteria-based ranking which is proposed in one of the recent state-of-the-art channel pruning techniques for computer vision applications. On the other hand, our proposed hybrid pruning technique uses a new approach that combines unstructured and structured ranking methods to achieve structured compression.

To make sure the network works flawlessly with the updated output shape of different layers during and after compression, the kernel sizes of the pooling layers are adjusted accordingly. Furthermore, after flattening the *Avgpool1* output, when the flattened vector has elements less than the number of classes (i.e., 50), the number of input neurons of the *Dense1* layer is dynamically adjusted so that it can handle the incoming input data to produce 50 output elements for the softmax output layer. During this process, if the kernel size of a pooling layer becomes (1,1), we remove that layer from the network.

During the compression process, we remove 80% and 85% of the channels in two runs, respectively, from the original network. These amounts are selected to obtain a model small enough for our target MCUs.

*4.1. Channel pruning using magnitude-based ranking*

In this method, the channels are ranked using their individual sum of absolute weights (L1 Norm) followed by layer-wise L2 normalization. Many approaches remove all the channels having a sum below a predefined threshold [30,32,33]. However, we remove the channels having the lowest sum iteratively (one at a time) and retrain the network to recover the loss until the network reaches the target size required to be fitted in the MCU. Let $Z$ be the a vector containing the absolute sum of the channels layer-wise, $l$ be the layer index and $i$ be the channel index. Then, an iteration of this pruning process is defined by:

$$Z_l = \sum |z_l^{(i)}|$$
$$Z_{rank} = sort(\Omega(Z_l))$$
$$Z_{prune} = argmin(Z_{rank}) \quad (10)$$

*4.2. Channel pruning using taylor criteria-based ranking*

This is an iterative channel pruning technique recently proposed by Molchanov et al. [32]. In this approach, the channels are ranked using Taylor criteria and the lowest ranked channel is pruned in a pruning iteration. More specifically, the ranking of filters is calculated by conducting a forward pass of the trained network for the whole dataset and observing the change to the cost function. The least affected channel after layer-wise L2 normalization is ranked highest to be pruned. The iterative pruning and retraining process continues until the model reaches the target size. An iteration of this pruning process is defined by:

$$Z_l = z_l + \overline{(z_l * \Delta_l)}$$
$$Z_{rank} = sort(\Omega(Z_l))$$
$$Z_{prune} = argmin(Z_{rank}) \quad (11)$$

where, $z_l$ holds the activations of the layers of the network and $\Delta_l$ is the gradient.

*4.3. Proposed hybrid pruning approach*

This pruning technique uses a new approach that combines both the unstructured and structured ranking methods to prune weights and channels from a network. In the first step, it prunes unimportant weights by zeroing them out from the network using unstructured global weight ranking methods, for example L0 Norm. In the second step, the model is further pruned through structured pruning. As we apply structured pruning on the weight-pruned network, we are not producing sparse matrices, which are generally not supported on embedded devices. In this step, the focus is to prune the important weights through channel pruning.

**Table 7**
Sparsifying the Weights in ACDNet.

| % Sparse | Original Accuracy | Final Accuracy |
|---|---|---|
| 85% | 91.00% | 90.50% |
| 90% | 91.00% | 89.25% |
| 95% | 91.00% | 91.00% |
| 98% | 91.00% | 88.75% |

The channels are first ranked by using structured channel ranking methods, such as magnitude-based ranking and Taylor criteria-based ranking. Then the lowest ranked channel is removed from the network and retrained the network (we term it as 'fine-tuning') for few epochs to recover the loss. This pruning and fine-tuning is iterated until the model reaches the target size. Then the existing weights of the resulting model is re-initialized and trained as a fresh network (we term it as 'scratch-training') instead of retraining the resulting model for higher classification accuracy (we term this as 're-training'). We present this method in the following form.

*4.4. Experimental results*

We use the same experimental setup that has been discussed in Section 3.2 for this experimental study. ESC-50 (fold-4) is used to fine-tune the network during the compression process. The resulting network is cross-validated on all the four datasets and the results are reported.

*4.4.1. Compressing ACDNet*
To compare our proposed method, at first we compress ACDNet using magnitude based pruning and taylor pruning. Then we experiment with two different combinations of the methods for testing our proposed iterative hybrid pruning approach. The first one consists of L0 Norm followed by channel pruning using magnitude-based ranking and the second one of L0 Norm followed by channel pruning using Taylor criteria-based ranking. Finally, we compare the results.

In this process, we first sparsify ACDNet by zeroing out 85%, 90%, 95% and 98% weights using L0 norm. We opted for the 95% sparsed network since it produces the same classification accuracy as the original ACDNet (see Table 7).

The sparsified model is further pruned through channel pruning. We prune 80% and 85% channels using all the discussed methods and present the results in Tables 8 and 9. Models 1-2, 3–4, 5–6 and 7–8 in both the tables are the resulting models of magnitude based pruning, taylor pruning and the two combinations of our proposed hybrid pruning respectively. All models are iteratively pruned and fine-tuned. The column *Fine-tuning Accuracy* shows the accuracy obtained at the end of the iterative pruning and fine-tuning process.

From the tables, we observe that pruning and fine-tuning does not recover the loss in accuracy as hoped for. To achieve the best accuracy, we therefore conduct further full re-training of the networks with their existing weights, scratch-training by re-initializing their weights. We use the base network's training settings for re-training and scratch training. The accuracy of these different training processes are reported in *Re-training Accuracy* and *Scratch-training Accuracy* columns respectively. In addition to re-training and scratch-training, we also train the re-initialized fresh networks using knowledge distillation [34] and the result is presented in Table 10.

Tables 8 and 9 show that the compressed networks with scratch-training produce better classification accuracy. To further justify this argument, we check the results obtained by training the models using knowledge distillation. The current weights of the compressed network are re-initialized and trained as a new network with distilled loss. In this process, the best performing 80% compressed model (i.e., Model 8 from Table 8) is used as the student network while the trained base ACDNet model acts as the teacher network. We apply three different custom loss functions (see Eq. 12, 13, and 14 using CrossEntropy and KLDivLoss loss, where the student output and the teacher output are denoted by SO and TO, respectively. We use $\alpha = 0.1$, $\beta = 1 - \alpha$ and Entropy (T) $\in [1-5]$. Table 10 shows the results of this training approach.

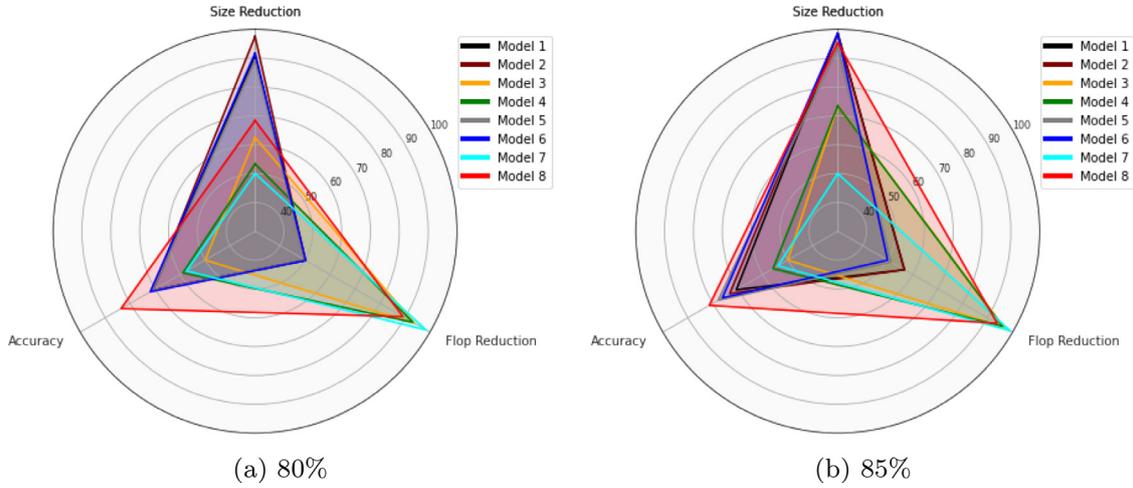

**Fig. 2.** (a) Comparison of 80% pruned models from Table 8. (b) Comparison of 85% pruned models from Table 9. For visualization, the variables are linearly transformed to range [50 - max(variable)].

$$L1 = CELoss(SO/T, TO/T) * \beta * T^2 + CELoss(SO, Target) * \alpha \quad (12)$$

$$L2 = KLDivLoss(SO/T, TO/T) * \beta * T^2 + CELoss(SO, Target) * \alpha \quad (13)$$

$$L3 = KLDivLoss(SO/T, TO/T) * \beta * T^2 + KLDivLoss(SO, Target) * \alpha \quad (14)$$

From Tables 8, 9 and 10 it is evident that our proposed hybrid pruning approach including scratch-training outperforms other approaches in terms of size and FLOPs reduction, and prediction accuracy. Note that for all the experiments, we have performed 5-fold CV.

*4.4.2. Selecting the compressed network*

We choose the best compressed network depending on three factors: compression, FLOP reduction and accuracy. Considering the result on all three measurements for 80% pruning shown in Table 8 and Fig. 2, we select Model 8, obtained by our hybrid pruning technique. For this model, our hybrid pruning technique achieves 97.22% model size reduction, 97.28% FLOP reduction (Table 8) and 83.65% CV accuracy on ESC-50 dataset (Table 14). This is still very close to the state of the art and significantly higher than human accuracy (81.30%). We term this network Micro-ACDNet as it only retrains 20% filters/channels.

With 85% pruning, our hybrid technique produces an even smaller version of Model 8 with 312 (15% retained) and only 240kB model size, which we term Macro-ACDNet (Fig. 2b). We select Micro-ACDNet over this macro version for three reasons. Firstly, Micro-ACDNet delivers significantly higher accuracy. Secondly, its model size of approximately 500kB already fits into typical MCU flash sizes and can be further decreased fourfold by 8-bit quantization to a size below the memory limits of even small microcontrollers. Thirdly, both models require approximately the same amount of FLOPs. Table 11 presents the architecture of Micro-ACDNet.

**Method:** Hybrid Pruning
**Input**: Trained Model, Target (size/percentage)
**Output**: Compressed Model
**Data**: Training Set, Test Set
**Step 1:**
 - Sparsify the trained model
**Step 2:**
 **while** model_size > Target **do**
  - A full epoch forward pass
  - Calculate the effect in gradient
  - Update activation
  - Layer-wise normalization
  - Global ranking of channels
  - Prune lowest ranked channel
  - Fine-tune for 2 epochs
**Step 3:**
 - Re-initialize the weights
 - Scratch-training: Train the model from scratch

*4.4.3. Compressing other networks*

Now, we perform the same methodology that is used to derive Micro-ACDNet to find a micro version of AclNet. We do this to verify whether our proposed compression approach generalizes well for other networks. We name the resulting compressed AclNet as Micro-AclNet. The output filter configurations of the 12 convolution layers of Micro-AclNet are 7, 26, 10, 20, 26, 41, 22, 48, 55, 64, 58 and 50. Micro-AclNet also produces comparable classification accuracy on ESC-50 while comparing it with its base network accuracy. Table 12 clearly shows that the method works as expected in compressing other networks.

*4.4.4. Compressed networks on different datasets*

Now we test and cross validate Micro-ACDNet and Micro-AclNet on all the four datasets used for this research. Since we have already established that training from scratch works better and Micro-ACDNet achieves the highest prediction accuracy, we conduct scratch-training for all the networks, 5-fold (ESC-10&50) and 10-fold (US8K) CV over five independent runs. Table 13 presents their classification accuracy on different datasets. The results in the table clearly indicates that the compressed models generalize well on all the four datasets. Their classification accuracy is very close to their base models classification accuracy.

**Table 8**
Models found after 80% channel pruning using magnitude-based ranking, Taylor criteria-based ranking and our hybrid pruning approach.

| Model No. | Pruning Method | SFEB | TFEB | Parameters | | | FLOPs | | Fine-tuning | CV Accuracy% | |
|---|---|---|---|---|---|---|---|---|---|---|---|
| | | | | Count(M) | Size(MB) | Reduced% | Count(M) | Reduced% | Accuracy% | Re-training | Scratch-training |
| 1 | Magnitude | | ✓ | 0.119 | 0.45 | **97.49** | 36.94 | 93.22 | 15.25 | 81.80 | 82.30 |
| 2 | Magnitude | ✓ | ✓ | 0.119 | 0.45 | 97.57 | 36.94 | 93.22 | 15.25 | 82.45 | 82.30 |
| 3 | Taylor | | ✓ | 0.135 | 0.52 | 97.15 | 11.03 | 97.97 | 18.00 | 77.45 | 80.05 |
| 4 | Taylor | ✓ | ✓ | 0.140 | 0.53 | 97.04 | 11.40 | 97.91 | 6.50 | 78.40 | 81.00 |
| 5 | Hybrid-Magnitude | | ✓ | 0.120 | 0.46 | 97.47 | 36.96 | 93.21 | 12.00 | 81.85 | 82.30 |
| 6 | Hybrid-Magnitude | ✓ | ✓ | 0.118 | 0.45 | 97.50 | 36.81 | 93.24 | 12.50 | 82.10 | 82.40 |
| 7 | Hybrid-Taylor | | ✓ | 0.142 | 0.54 | 97.00 | 8.22 | **98.49** | 23.75 | 80.30 | 80.85 |
| 8 | Hybrid-Taylor | ✓ | ✓ | 0.131 | 0.50 | 97.22 | 14.82 | 97.28 | 48.50 | 81.30 | **83.65** |

**Table 9**
Models found after 85% channel pruning using magnitude-based ranking, Taylor criteria-based ranking and our hybrid pruning approach.

| Model No. | Pruning Method | SFEB | TFEB | Parameters | | | FLOPs | | Fine-Tuned | CV Accuracy% | |
|---|---|---|---|---|---|---|---|---|---|---|---|
| | | | | Count(M) | Size(MB) | Reduced% | Count(M) | Reduced% | Accuracy% | Re-Trained | Scratch-Training |
| 1 | Magnitude | | ✓ | 0.065 | 0.25 | **98.63** | 22.61 | 95.85 | 3.00 | 79.65 | 79.65 |
| 2 | Magnitude | ✓ | ✓ | 0.065 | 0.25 | **98.63** | 22.61 | 95.85 | 3.00 | 79.20 | 80.15 |
| 3 | Taylor | | ✓ | 0.072 | 0.27 | 98.48 | 6.09 | 98.88 | 13.25 | 72.95 | 76.60 |
| 4 | Taylor | ✓ | ✓ | 0.072 | 0.28 | 98.48 | 6.31 | 98.84 | 9.50 | 73.80 | 77.50 |
| 5 | Hybrid-Magnitude | | ✓ | 0.066 | 0.25 | 98.60 | 24.82 | 95.44 | 10.75 | 78.85 | 80.85 |
| 6 | Hybrid-Magnitude | ✓ | ✓ | 0.065 | 0.25 | **98.63** | 25.39 | 95.34 | 6.75 | 79.90 | 80.60 |
| 7 | Hybrid-Taylor | | ✓ | 0.079 | 0.30 | 98.34 | 5.01 | **99.08** | 15.50 | 74.10 | 77.20 |
| 8 | Hybrid-Taylor | ✓ | ✓ | 0.066 | 0.25 | 98.61 | 10.54 | 98.68 | 37.00 | 76.90 | **81.40** |

**Table 10**
Performance of Micro-ACDNet when trained using Knowledge Distillation.

| Loss | 5-Fold CV Accuracy% | | | | |
|---|---|---|---|---|---|
| | T=1 | T=2 | T=3 | T=4 | T=5 |
| L1 | 81.45 | 80.15 | 78.75 | 78.45 | 78.35 |
| L2 | 81.70 | 81.90 | 81.40 | 81.65 | 81.95 |
| L3 | 81.55 | **81.95** | 81.75 | 81.25 | 81.85 |

**Table 11**
Micro-ACDNet architecture for input length 30,225 (approximately 1.51s audio @ 20kHz).

| Layers | Kernel Size | Stride | Filters | Output Shape |
|---|---|---|---|---|
| Input | | | | (1, 1, 30225) |
| conv1 | (1, 9) | (1, 2) | 7 | (7, 1, 15109) |
| conv2 | (1, 5) | (1, 2) | 20 | (20, 1, 7553) |
| Maxpool1 | (1, 50) | (1, 50) | | (20, 1, 151) |
| swapaxes | | | | (1, 20, 151) |
| conv3 | (3, 3) | (1, 1) | 10 | (10, 32, 151) |
| Maxpool2 | (2, 2) | (2, 2) | | (10, 16, 75) |
| conv4 | (3, 3) | (1, 1) | 14 | (14, 16, 75) |
| conv5 | (3, 3) | (1, 1) | 22 | (22, 16, 75) |
| Maxpool3 | (2, 2) | (2, 2) | | (22, 8, 37) |
| conv6 | (3, 3) | (1, 1) | 31 | (31, 8, 37) |
| conv7 | (3, 3) | (1, 1) | 35 | (35, 8, 37) |
| Maxpool4 | (2, 2) | (2, 2) | | (35, 4, 18) |
| conv8 | (3, 3) | (1, 1) | 41 | (41, 4, 18) |
| conv9 | (3, 3) | (1, 1) | 51 | (51, 4, 18) |
| Maxpool5 | (2, 2) | (2, 2) | | (51, 2, 9) |
| conv10 | (3, 3) | (1, 1) | 67 | (67, 2, 9) |
| conv11 | (3, 3) | (1, 1) | 69 | (69, 2, 9) |
| Maxpool6 | (2, 2) | (2, 2) | | (69, 1, 4) |
| Dropout (0.2) | | | | |
| conv12 | (1, 1) | (1, 1) | 48 | (48, 1, 4) |
| Avgpool1 | (1, 4) | (1, 4) | | (48, 1, 1) |
| Flatten | | | | (48) |
| Dense1 | | | | (50) |
| Softmax | | | | (50) |

#### 4.4.5. Comparing micro-ACDNet with other networks

Now that we have the resulting compressed model in our hand, we can compare it with Edge-L[3], the only existing edge audio architecture. Table 14 shows our results in comparison to Edge-L[3].

#### 4.4.6. Experimental findings

Tables 8 and 9 provide the experimental results for the different pruning approaches tested on ACDNet. In Table 8 we see that our hybrid approach (Models 5-8) produces the best prediction accuracy and FLOP reduction for a model size reduction that fits the target specifications (Model 8).

Pruning channels from initial convolution layers leads to more size and FLOPs reduction and causes little accuracy loss, thus making it a suitable approach for extremely resource-constrained MCUs.

From the experiments, we observe that *pruning and fine-tuning* (each iteration consists of pruning one channel and retraining the resulting network for two epochs) is not enough to achieve comparable prediction accuracy, which is contrary to earlier findings in the literature, e.g., in Molchanov et al. [32]. The experimental results (Table 8 and 9) show that the final compressed network requires full *re-training* or *scratch-training* to produce comparable prediction accuracy.

Our experiments furthermore suggest, supported by earlier work by Crowley et al. [81] and Liu et al. [82], that pruning or compression should be considered as an efficient DL model architecture search method.

**Table 12**
ACDNet vs AclNet performance after 80% compression.

| Network | Pruning Method | SFEB | TFEB | Size | | | FLOPs | | | Scratch-Training |
|---|---|---|---|---|---|---|---|---|---|---|
| | | | | Base (MB) | Compressed (MB) | Reduced% | Base (M) | Compressed (M) | Reduced% | Accuracy (%) |
| ACDNet | Hybrid-Taylor | ✓ | ✓ | 18.06 | 0.50 | 97.22 | 544.42 | 14.82 | 97.28 | **83.65** |
| AclNet | Hybrid-Taylor | ✓ | ✓ | 40.54 | 0.50 | 98.79 | 1806.57 | 21.50 | 98.81 | 80.05 |

**Table 13**
ACDNet, AclNet, Micro-ACDNet and Micro-AclNet performance on different datasets.

| Accuracy (%) → | Networks | | | |
|---|---|---|---|---|
| Datasets ↓ | ACDNet | AclNet | Micro-ACDNet | Micro-AclNet |
| ESC-10 | 96.75 | 95.75 | 96.25 | 94.00 |
| ESC-50 | 87.10 | 85.65 | 83.65 | 80.05 |
| US8K | 84.45 | 79.17 | 78.28 | 75.80 |
| AE | 92.57 | 90.15 | 89.69 | 87.51 |

**Table 14**
Parameters, size and computation requirements for ACDNet and Micro-ACDNet for approximately 1.51s audio @ 20kHz.

| Networks | #Filters | Params (M) | Size (MB) | FLOPs (B) | Accuracy (%) |
|---|---|---|---|---|---|
| ACDNet | 2074 | 4.74 | 18.06 | 0.54 | 87.10 ± 0.02 |
| Edge-L$^3$ [63] | 1920 | 0.213 | 0.814 | - | 73.75 |
| Micro-ACDNet | 415 | 0.131 | 0.50 | 0.0148 | 83.65 |

## 5. Deployment in MCU (edge-AI device)

Our network has been fully deployed on an off-the-shelf MCU. The most limiting factor for MCU deployment is memory requirements. It is important to distinguish between two memory types: (1) Flash memory, which is not suitable for fast, frequent write access and is used to hold the (fixed) model parameters and (2) SRAM, which is used to store inputs and the results of intermediate calculations, i.e. activation values. As mentioned above, typical parameters for widely used highly power-efficient MCUs are less than 1MB of Flash and less than 512kB of SRAM.

Most off-the-shelf MCUs have either no audio capability or very low quality audio on-board. One could, in principle, add additional audio hardware to achieve better quality. However, this would increase power consumption. An exception is the Sony Spresense [83] which provides high-quality audio input and processing on-board. We have selected this unit for these reasons.

While the Spresense is a highly power-efficient device, it has somewhat more generous specifications than the most common MCUs, providing 1.5MB SRAM. It is thus important to note that we are not actually making use of this additional memory. Our final deployed model requires 303kB SRAM for intermediate calculations, which is well below the target. This can be further reduced to below 200kB using a hand-optimised implementation as detailed below. While we have not yet taken this last step in a physical deployment, the implication is that Micro-ACDNet can fit on even smaller MCUs, such as those based on the extremely popular Nordic nRF52840 SoC, which offers 256kB SRAM and 1MB Flash. Flash memory is not a limiting constraint, since Micro-ACDNet requires only 500kB of model storage (Table 14).

We picked Tensorflow Lite Micro over the other platform-independent DNN software frameworks for small-device deployment (e.g., PyTorch Mobile, Tensorflow Lite Micro) because it is the most frequently used and the only one that allows us to implement ACDNet directly. ACDNet's *transpose* layer is not currently supported by PyTorch Mobile.

To achieve the required model size, we need to quantize ACDNet. The present paper is not concerned with quantization itself, so that we simply apply the quantization methods directly available in Tensorflow Lite Micro. Quantization, unfortunately, reduces the model accuracy noticeably and the micro version of Tensorflow Lite does not seem to provide the best results here. We used 8-bit post-training quantization which reduces the accuracy to 71.00%.

**Table 15**
Prediction accuracy on ESC-50 after quantization.

| Network | Library | Quantized Size | Quantized Accuracy (%) |
|---|---|---|---|
| Micro-ACDNet | Pytorch | 157kB | 81.50% |
| Micro-ACDNet | TF Lite Micro | 153kB | 71.00% |

To confirm that better results are possible, we also tested alternative frameworks. We found that Pytorch achieved the highest accuracy with 81.50% for an 8-bit quantized model of equivalent size (see Table 15). While our actual deployment uses Tensorflow Lite Micro for implementation-related reasons, we conclude that an alternative version of ACDNet that achieves 81.5% accuracy (i.e. above human performance) can be deployed on a standard MCU of the same size.

The required working memory of 303kB SRAM could be reduced further. The current requirement results from essentially computing tensors layer by layer, keeping one intermediate layer in memory at a time. This can be optimised by re-grouping computations. The bottleneck for working memory is *conv2*. However, this is immediately followed by a non-overlapping max pooling layer. Therefore, each instance of *Maxpool2* can be immediately computed and the corresponding slice of output from *conv2* can immediately be discarded after this. In this way, at any point of time, the most we need to keep in memory is the output of *conv1*, a 110 units wide segment of *conv2* and the complete output of *Maxpool1*. This modification would allow us to reduce the working memory bottleneck to 141,636 bytes. While this may be detrimental to GPU speed-ups, it does not impact on an MCU implementation. However, this cannot be done in Tensorflow and would require a manual implementation.

## 6. Future work and conclusions

We have presented the first implementation of a 50 class audio classifier that achieves high accuracy, yet is small enough to fit on MCUs commonly used in energy-efficient internet-of-things devices. We constructed a full size network that sets a new standard for all the four datasets ESC-10, ESC-50, US8K and AE benchmarks for raw audio classification and compressed this for MCU deployment. While limitations of the programming environment have restricted the accuracy of our current test deployment on a physical MCU, we have conclusively shown that 81.50% accuracy is achievable on such a resource-impoverished device, close to the state-of-the-art and above human performance. This brings our goal of continuous, autonomous animal monitoring into immediate reach and should open new horizons for many other audio applications on the internet-of-things.

Our deployment of a DNN originally not designed for MCUs has shown that structured pruning achieves the best results for this task. It has also highlighted the fact that machine learning on the edge, still very much cutting edge, is not yet sufficiently supported by standard frameworks. We particularly expect that future frameworks will provide improved quantization support.

Some next steps immediately arise. Firstly, it is likely that the performance can be improved by using quantization-aware training and pruning. Secondly, we would like to try the SpArSe approach for further optimisations now that we have developed Micro-ACDNet as a suitable starting point for its optimisations.

We believe it to be of particular importance that we have constructed and used a generic pipeline to derive an MCU implementation from a standard DNN. This opens up the same opportunities for a wide range of applications and we are confident that we will be able to transfer our approach to other domains.

**Declaration of Competing Interest**

The authors declare no conflict of interest.

**Acknowledgements**

This work was supported in part by the Australian Research Council under grant DE190100045.

We thank Lin Schwartzkopf and Slade Allen-Ankins of James Cook University in Townsville for the provision of the real-world frog dataset and Akhter Hossain of Sanofi US for advice on statistical significance testing.

**Md Mohaimenuzzaman** is a PhD candidate in the Department of Data Science and AI, Faculty of Information Technology at Monash University, Australia. Before starting his PhD, he worked as a Software Engineer for about a decade. He received his bachelors and masters degrees in Computer Science and Engineering in 2007 and 2013.

**Christoph Bergmeir** is a Senior Lecturer in Data Science and AI, and a 2019 ARC DECRA Fellow in the Department of Data Science and AI at Monash University. Christoph holds a PhD in Computer Science from the University of Granada, Spain, and an M.Sc. degree in Computer Science from the University of Ulm, Germany.

**Ian Thomas West** (BCSE Hons, Monash University, 2002) the Director of Factorem Pty. Ltd. He has worked extensively in law enforcement, emergency services monitoring, and manufacturing fields and is currently developing micro frameworks for distributed autonomous decision making and machine learning on embedded devices.

**Bernd Meyer** is a Professor in the Department of Data Science and AI, Faculty of Information Technology at Monash University, Australia. He received his PhD in computer science in 1994. Bernd develops mathematical and computational models to explain the collective behavior of social insects. He also works on AI-based methods for ecosystem monitoring.